\documentclass[aps,prd,twocolumn,floatfix,nofootinbib,superscriptaddress]{revtex4}

\usepackage{amsmath}
\usepackage{amssymb}
\usepackage{amsthm}
\usepackage{bbm}
\usepackage{bm}
\usepackage{dcolumn}
\usepackage{epsfig}
\usepackage{graphicx}
\usepackage{longtable}
\usepackage{color}
\usepackage{epstopdf}
\usepackage{nicefrac}
\usepackage{tabularx}

\usepackage{tabularx}
\newcolumntype{L}[1]{>{\raggedright\arraybackslash}p{#1}}
\newcolumntype{C}[1]{>{\centering\arraybackslash}p{#1}}
\newcolumntype{R}[1]{>{\raggedleft\arraybackslash}p{#1}}

\usepackage[colorlinks=true, pdfstartview=FitV, linkcolor=purple, citecolor= purple,urlcolor=blue]{hyperref}
\definecolor{darkgreen}{rgb}{0,0.5,0}
\definecolor{purple}{rgb}{0.5,0,0.5}
\definecolor{nblue}{rgb}{0.0,0.0,0.50}
\definecolor{scarlet}{rgb}{1.0,0.2,0}

\begin{document}

\title{Gauge dependence of the quark gap equation: an exploratory study}

\author{Jos\'e Roberto Lessa}
\affiliation{Laborat\'orio de F\'isica Te\'orica e Computacional, Universidade Cidade de S\~ao Paulo, Rua Galv\~ao Bueno 868, S\~ao Paulo, S\~ao Paulo 01506-000, Brazil}
\email{ze$_$roberto$_$lessa@hotmail.com}

\author{Fernando E. Serna}
\email{fernando.serna@unisucrevirtual.edu.co}
\affiliation{Laborat\'orio de F\'isica Te\'orica e Computacional, Universidade Cidade de S\~ao Paulo, Rua Galv\~ao Bueno 868, S\~ao Paulo, S\~ao Paulo 01506-000, Brazil}
\affiliation{Departamento de F\'isica, Universidad de Sucre, Carrera 28 No.~5-267, Barrio Puerta Roja, Sincelejo 700001,  Colombia}

\author{Bruno El-Bennich}
\email{bruno.bennich@cruzeirodosul.edu.br}
\affiliation{Laborat\'orio de F\'isica Te\'orica e Computacional, Universidade Cidade de S\~ao Paulo, Rua Galv\~ao Bueno 868, S\~ao Paulo, S\~ao Paulo 01506-000, Brazil}
\affiliation{Departamento de F\'isica, ICAFQ, Universidade Federal de S\~ao Paulo, Diadema, S\~ao Paulo 09913-030, Brazil}

\author{Adnan Bashir}
\email{adnan.bashir@umich.mx;abashir@jlab.org}
\affiliation{Instituto de F\'isica y Matem\'aticas, Universidad Michoacana de San Nicol\'as de Hidalgo, Morelia, Michoac\'an 58040, Mexico}
\affiliation{Theory Center, Jefferson Lab, Newport News, VA 23606, USA}

\author{Orlando Oliveira}
\email{orlando@teor.fis.uc.pt}
\affiliation{CFisUC, Department of Physics, University of Coimbra, 3004 516 Coimbra, Portugal}


\begin{abstract}
We study the gauge dependence of the quark propagator in quantum chromodynamics (QCD) by solving the gap equation with a nonperturbative quark-gluon vertex which is 
constrained by longitudinal and transverse Slavnov-Taylor identities, the discrete charge conjugation and parity symmetries and which is free of kinematic 
singularities in the limit of equal incoming and outgoing quark momenta. We employ gluon propagators in renormalizable $R_\xi$ gauges obtained in lattice QCD studies. 
We report the dependence of the nonperturbative quark propagator on the gauge parameter, in particular we observe an increase, proportional to the gauge-fixing parameter, 
of the mass function in the infrared domain, whereas the wave renormalization decreases within the range $0 \leq \xi \leq 1$ considered here.  The chiral quark condensate 
reveals a mild gauge dependence in the region of $\xi$ investigated. We comment on how to build and improve upon this exploratory study in future in 
conjunction with generalized gauge covariance relations for QCD.
\end{abstract}

%

\date{\today}

\maketitle



\section{Introduction}

Gauge symmetry and its innumerable consequences have played a fundamental role in the development of modern quantum field theory during the past century. Amongst 
its celebrated implications, the Ward-Takahashi identities (WTI)~\cite{Ward:1950xp,Takahashi:1957xn} in quantum electrodynamics (QED) and the corresponding Slavnov-Taylor 
identities (STI)~\cite{Slavnov:1972fg,Taylor:1971ff} of quantum chromodynamics (QCD) relate different $n$-point Green functions to each other. In  particular, the WTI and the 
STI connecting the {\em divergence} of the fermion-boson vertex to the fermion propagator help us identify the {\em longitudinal} part~\cite{Ball:1980ay,Aguilar:2016lbe,
Bermudez:2017bpx} of this three-point vertex, whether it be the quark-photon or the quark-gluon vertex. These are exact nonperturbative relations which are observed 
order by order in perturbation theory. There exists a plethora of work within the nonperturbative exploration of Dyson-Schwinger equations (DSEs) which incorporates these 
identities in studying dynamical chiral symmetry breaking (DCSB) via the electron/quark gap equation. 

Whereas the usual WTI or STI relates the divergence of the three-point fermion-boson vertex to the inverse fermion propagator, there exist transverse Takahashi identities (TTI) 
and transverse Slavnov-Taylor identities (TSTI) which play a similar role for the curl of the fermion-boson vertex~\cite{Takahashi:1985yz,Kondo:1996xn,He:2000we,He:2006my,
He:2007zza}. The TTI and TSTI are richer and more complicated in their structure, and they shed light on the {\em transverse} part of the fermion-boson vertex. They have been 
employed to compute the critical coupling and study its gauge independence, as well as the quark condensate and pion decay constant~\cite{He:2009sj,Qin:2013mta,
Qin:2014vya,Albino:2018ncl,Albino:2021rvj}. 

While the longitudinal and transverse gauge identities relate different $n$-point Green functions with each other, another important consequence of gauge covariance
in QED are the Landau-Khalatnikov-Fradkin transformations (LKFTs) which describe how the individual Green functions respond to an arbitrary gauge 
transformation~\cite{Landau:1955zz,Fradkin:1955jr}. The LKFTs are a well-defined set of transformations which leave the DSEs and related WTIs of the fermion-boson 
vertex form invariant. While the STIs are the QCD generalization of the WTIs, the equivalent generalization of the LKFTs has only recently been derived by two groups 
employing two different methods: (i) ABG~\cite{Aslam:2015nia} through direct generalization of the method employed by Landau and Khalatnikov  and (ii) MDSDB~\cite{DeMeerleer:2018txc,PhysRevD.101.085005} based on the introduction of a gauge invariant transverse gauge field. Moreover, the Nielsen identities allow 
to study the variation of Green functions as derivatives with respect to the gauge parameter~\cite{PhysRevD.101.085005,NIELSEN1975173}. Nevertheless, even with 
these formal results available, starting from the knowledge of Green functions in a given gauge, their explicit extraction in another gauge remains a nontrivial problem 
in QCD even at one loop in perturbation theory. This is already the case for two-point Green functions, let alone their nonperturbative transformation and that of three- 
and higher $n$-point Green functions. 

These local transformations are in no simple manner amenable to straightforward comparisons of Green functions in different gauges, especially in momentum space, 
nor to explicitly prove the gauge invariance of physical observables. However, it has explicitly been shown in QED, and verified in QED3 by constructing a 
nonperturbative vertex~\cite{Bashir:2011vg}, studying gauge-covariance relations~\cite{Bashir:1999bd,Bashir:2000rv} and generating numerical solutions of the 
electron gap equation, that the electron condensate is manifestly gauge independent once its propagator is correctly LKF transformed from a given covariant gauge 
to other covariant gauges~\cite{Bashir:2004yt,Bashir:2005wt}. Moreover, the ABG and the MDSDB generalizations of LKFT in QCD confirm the formal gauge-invariance 
of the quark  condensate after the quark propagator has been adequately gauge transformed. 

We are not able to directly verify this result by gauge transforming every Green function computed in Landau gauge to repeat the calculation in any other gauge---it
remains a prohibitively difficult task for the time being. We thus follow a more modest approach to study the gauge dependence of the quark propagator in QCD and 
\emph{solve\/} the associated DSE, employing quenched gluon propagators in $R_\xi$ gauges from lattice QCD~\cite{Bicudo:2015rma}, and the dressed quark-gluon 
vertex constructed in Ref.~\cite{Albino:2021rvj}, invoking the STI and TSTI. Our approach not only  allows for a comparison of the DSE solutions in $R_\xi$ gauges,  
i.e., the mass and wave renormalization functions of the quark for gauge parameters in the range  $\xi \in [0,0.5]$, extrapolated to Feynman gauge, $\xi=1$, but also 
to compute the quark condensate as a function of the gauge parameter within this interval.  We emphasize that this is an initial, exploratory study which we expect to 
shed light on how far we are from obtaining the gauge independence of the condensate. This invariance would be obtained through formal incorporation of the local 
gauge transformations of every Green function in the gap equation, i.e., the gluon propagator, the quark-gluon vertex as well as the ghost propagator and the 
ghost-quark scattering kernel.

Within the limitations of this hybrid approach, we demonstrate how the numerical solutions of the quark propagator vary as function of the gauge parameter 
in $R_\xi$  gauges, and that this behavior leads to a slightly gauge-dependent quark condensate up to Feynman gauge within the error estimates involved.
Nonetheless, to our knowledge, we present the first DSE solutions for the quark obtained with all 12 vector structures of the nonperturbative quark-gluon vertex 
and the gluon and ghost propagators in covariant $R_\xi$  gauges. Our findings are encouraging and represent  an important first step as they incorporate the 
gauge covariance into the DSEs through the formal implementation of gauge identities for the Green functions. This is a necessary, though not sufficient requirement 
for calculating gauge-independent bound-state properties.

This article is organized as follows: in Sec.~\ref{secDSE} we describe our functional approach to QCD and discuss the quark-gluon vertex, the lattice-extracted gluon 
and ghost propagators and the quark-ghost form factor in different gauges. In Sec.~\ref{sec3}, we solve the quark gap equation and obtain the mass and wave 
renormalization functions in different gauges and calculate the quark condensate. Final remarks are given in Sect.~\ref{sec4}.


\section{Dyson-Schwinger equation in \textit{R}$_\mathbf{\xi}$ gauges}
\label{secDSE}

The DSEs are the relativistic equations of motion in quantum field theory, see e.g. Ref.~\cite{Bashir:2012fs} for a review.
For a given flavor and in $R_\xi$ gauge, the DSE of the inverse quark propagator in Euclidean space reads,
\begin{align}
 \label{DSEquark} 
  S^{-1}_\xi(p) & = \, Z_2 \,  i\, \gamma\cdot p + Z_4 \, m(\mu)  \nonumber   \\ 
             + \, Z_1 & 4\pi \alpha_s^\xi \! \int^\Lambda\!\!  \frac{d^4k}{(2\pi)^4}\  \Delta^{ab}_{\mu\nu} (q)\, \gamma_\mu t^a\,  S_\xi (k) \,\Gamma_\nu^{b\xi} (k,p) \, ,                       
\end{align}
where $m(\mu)$ is the renormalized current-quark mass, \linebreak $Z_1(\mu,\Lambda)$,   $Z_2(\mu,\Lambda)$ and $Z_4(\mu,\Lambda)$ are the vertex,  wave  \linebreak function 
and mass renormalization constants, respectively, while $\Gamma^{a\xi}_\mu (k,p)=  \Gamma_\mu^\xi (k,p) t^a$ is the dressed quark-gluon vertex and 
$t^a = \lambda^a/2$  are the SU(3) group generators in the fundamental representation. The gluon propagator in $R_\xi$ gauge with momentum $q=k-p$,
\begin{equation}
\Delta^{ab}_{\mu\nu} (q ) =  \delta^{a b} \left(\delta_{\mu \nu}-\frac{q_{\mu} q_{\nu}}{q^{2}}\right ) \! \Delta_\xi (q^2)  
       + \delta^{ab}\, \xi\dfrac{q_{\mu}q_{\nu}}{q^4} \ 
 \label{gluonprop}   
\end{equation}    
is characterized by a nonperturbative transverse dressing function, $\Delta_\xi(\mu^2 ) =1/\mu^2$, and was studied with different 
lattice and functional approaches, e.g. Refs.~\cite{Bicudo:2015rma,Huber:2015ria,Napetschnig:2021ria}.

The most general Poincar\'e-covariant form of the solutions to Eq.~\eqref{DSEquark} is written in terms of covariant scalar and vector amplitudes:  
\begin{equation}
\label{DEsol}
   S_\xi (p)  =    \frac{1}{i \gamma \cdot p \,A_\xi (p^2)  + B_\xi ( p^2 ) } = \,  \frac{Z_\xi (p^2 )}{ i \gamma \cdot p + M_\xi ( p^2 )} \   .
\end{equation}
In the self-energy integral, $\Lambda$ is a Poincar\'e-invariant ultraviolet cutoff and $\mu$ is the renormalization scale imposed such that $\Lambda \gg \mu$. 
This scale is implicit in our convenient notation: $A(p^{2}) \equiv A(p^{2},\mu^2 )$ and $B(p^{2}) \equiv B(p^{2},\mu^2 )$, as is a flavor index
$f$ for these quantities and for all renormalization constants. The flavor- and gauge-dependent mass and wave renormalization functions are, respectively: 
\begin{align}
      M_\xi (p^2) & = B_\xi(p^{2},\mu^2 )/A_\xi(p^{2},\mu^2) \ , \\
     Z_\xi (p^2,\mu^2 ) & = 1/A_\xi (p^2,\mu^2) \ .
\end{align}
We choose the renormalization scale, $\mu = 4.3$~GeV, for a twofold reason:  \emph{i\/}) we consistently renormalize the DSE at the scale at which 
the transverse dressing function $\Delta_\xi (q^2)$ in $R_\xi$ gauge is renormalized~\cite{Bicudo:2015rma} and \emph{ii\/}) we can compare the dressed functions 
$M_\xi (p^2)$  and $Z_\xi (p^2,\Lambda^2 )$ with the solutions of lattice regularized  QCD at this scale~\cite{Bowman:2004jm,Oliveira:2018lln}. We therefore 
impose~\cite{Rojas:2013tza,Serna:2018dwk}:  $Z_\xi (\mu^2) =1$  and $M_\xi (\mu^2 ) \equiv  m(\mu^2) = 25$~MeV.

For the dressed quark-gluon vertex we employ the decomposition detailed, e.g., in Ref.~\cite{Albino:2018ncl},
\begin{align}
  \Gamma_{\mu}^\xi (k,p)  & = \Gamma^{L\xi}_{\mu}(k,p)  + \Gamma^{T\xi}_{\mu}(k,p)  \nonumber \\
                                          =   \sum_{i=1}^4 & \, \lambda_i^\xi (k,p) L_{\mu}^i(k,p) + \sum_{i=1}^8 \tau_i^\xi (k,p) T_{\mu}^i(k,p)  \ ,
\label{BallChiu}
\end{align}
 where the transverse vertex $\Gamma^{T\xi}_{\mu}(k,p) $ in Eq.~\eqref{BallChiu} is naturally defined by  $i\,  q \cdot \Gamma^{T\xi}  (k,p)  =  0$ .
The usual STI~\cite{Slavnov:1972fg,Taylor:1971ff} constrains the {\em longitudinal} vertex, $\Gamma^{L\xi}_{\mu}(k,p)$, to four independent structures 
but leaves the transverse components undetermined. The remaining eight tensor structures can be explored with TSTI derived from the 
symmetry transformation that involves the Lorentz transformation acting on the usual infinitesimal gauge transformation. 

The TSTI~\cite{He:2009sj} constrain but also couple the  vector and axial-vector vertices. However, as shown in Refs. \cite{Qin:2013mta,Albino:2018ncl},
these can be decoupled  and merely one identity for the vector vertex is sufficient to obtain analytic expressions for the transverse form factors $\tau_i^\xi (k,p)$. The full 
vertex is thus described by the form factors derived in Refs.~\cite{Albino:2021rvj,El-Bennich:2022obe},
\begin{eqnarray}
  \lambda_1^\xi (k, p)  & = & \tfrac{1}{2}\,  G(q^2) X_0^\xi  (q^2)\left[ A_\xi (k^2) + A_\xi (p^2)   \right] \ , 
 \label{lambda1QCD}   \\ [5pt] 
   \lambda_2^\xi (k, p)  & = &  G(q^2) X_0^\xi  (q^2)\, \frac{ A_\xi  (k^2) - A_\xi (p^2)}{k^2-p^2}  \ , 
 \label{lambda2QCD}   \\ [5pt] 
  \lambda_3^\xi (k, p)  & = &   G(q^2) X_0^\xi  (q^2)\, \frac{ B_\xi  (k^2) - B_\xi (p^2)}{k^2-p^2}  \ , 
 \label{lambda3QCD}   \\ [5pt] 
  \lambda_4^\xi (k, p)  & = &\ 0 \ ,
 \label{lambda4QCD}  
\end{eqnarray}
for the longitudinal part and by
\begin{eqnarray}
   \tau_1^\xi (k, p) & = & - \frac{ Y_{1} }{ 2 (k^{2} - p^{2}) \nabla(k,p) } \ ,
\label{tau1QCD}   \\ [5pt]   
   \tau_2^\xi (k, p) &=& - \frac{Y_{5} - 3 Y_{3}}{ 4 (k^{2} - p^{2}) \nabla(k,p) } \ ,
\label{tau2QCD} \\ [5pt]
   \tau_3^\xi (k, p) &=&  \frac{1}{2}\,  G(q^2) X_0^\xi  (q^2)\left[ \frac{ A_\xi  (k^2) - A_\xi (p^2)}{k^2-p^2}  \right] \nonumber \\
   & + & \frac{Y_{2}}{4\nabla(k,p)} 
   - \frac{ (k+p)^{2} (Y_{3} - Y_{5}) }{ 8(k^{2} - p^{2}) \nabla(k,p) } \ ,   
\label{tau3QCD}  \\  [5pt]
   \tau_4^\xi (k, p) &=&  - \frac{ 6 Y_{4} + Y_{6}^{A} }{8\nabla(k,p) } - \frac{(k+p)^{2} Y_{7}^{S} }{ 8(k^{2} - p^{2}) \nabla(k,p) } \  , 
\label{tau4QCD}    \\ [5pt]
 \tau_5^\xi (k, p) &=& -  G(q^2) X_0^\xi  (q^2)\left[ \frac{ B_\xi  (k^2) - B_\xi (p^2)}{k^2-p^2}  \right ]   \nonumber   \\
      & - &  \frac{2 Y_4 + Y_6^A}{2 (k^{2}-p^{2})} \ , 
\label{tau5QCD}  \\ [5pt]
   \tau_6^\xi (k, p) &=& \frac{(k-p)^{2} Y_{2} }{ 4 (k^{2} - p^{2}) \nabla(k,p) } -\frac{Y_3 - Y_5}{8\nabla(k,p) }  \ , 
\label{tau6QCD}   
\end{eqnarray}
\begin{eqnarray}
   \tau_7^\xi (k, p) &=& \frac{q^2 (6Y_4 +Y_6^A)}{4(k^2-p^2)\nabla(k,p)}   +\frac{Y_7^S}{4\nabla(k,p )} \ , 
\label{tau7QCD} \\ [5pt]
   \tau_8^\xi (k, p) &=&  - G(q^2) X_0^\xi  (q^2)\left[ \frac{ A_\xi  (k^2) - A_\xi (p^2)}{k^2-p^2}  \right]  \nonumber \\
   & - &  \frac{2 Y_8^A}{k^{2}-p^{2}}  \ , 
\label{tau8QCD}
\end{eqnarray}
for the transverse components, where the Gram determinant is defined as: $ \nabla (k,p) = k^{2} p^{2} - (k \cdot p)^{2} $.

\begin{table*}[t!]
\centering
\begin{tabular}{ C{1cm} |  C{2.3cm}| C{2.3cm} | C{2.3cm} | C{2.3cm}| C{2.3cm} | C{1.5cm} } \hline \hline
  $\xi$ &  $Z$  &  $M_0^2$  [GeV$^2$] &  $M_1^2$ [GeV$^2$]  & $M_2^2$  [GeV$^2$]  & $M_3^4$   [GeV$^4$] &  $\chi^2$/dof  \\ \hline
  0.0 &  $8.407  \pm 0.031$  &   $0.071\pm 0.023$  &  $2.783\pm 0.128$  &  $0.532\pm 0.055$  &  $0.380\pm 0.012$ & 1.209 \\  
  0.1 &  $8.321  \pm 0.023$  &   $0.071\pm 0.018$  &  $2.712\pm 0.099$  &  $0.517\pm 0.043$  &  $0.374\pm 0.009$ & 0.734 \\  
  0.2 &  $8.238 \pm 0.018$   &   $0.074\pm 0.015$  &  $2.677\pm 0.076$  &  $0.515\pm 0.033$  &  $0.371\pm 0.007$ & 0.421 \\
  0.3 &  $8.160  \pm 0.012$   &   $0.069\pm 0.010$  &  $2.617\pm 0.051$ &  $0.499\pm 0.023$  &  $0.367\pm 0.005$ & 0.205 \\ 
  0.4 &  $8.072  \pm 0.011$   &   $0.076\pm 0.010$  &  $2.595\pm 0.046$ &  $0.499\pm 0.020$  &  $0.363\pm 0.004$ & 0.178 \\  
  0.5 &  $7.996 \pm 0.009$   &   $0.075\pm 0.009$  &  $2.570\pm 0.041$  &  $0.500\pm 0.018$  &  $0.362\pm 0.004$  & 0.162 \\ \hline
  0.7 &  $7.831$   &   $0.077 $  &  $2.488$  &  $ 0.487$  &  $0.354$ &  ---  \\
  1.0 &  $7.585$   &   $0.080 $  &  $2.378$  &  $0.473$  &  $0.344$ &  --- \\ \hline \hline 
\end{tabular}
\caption{Parameters of the gluon-dressing parametrization~\eqref{gluonparam} as a function of the gauge parameter $\xi$. Note that the bare gluon propagators 
             of Ref.~\cite{Bicudo:2015rma} were fitted which must be renormalized as $\Delta_\xi(\mu^2 ) =1/\mu^2$. The values $\xi=0.7$ and $\xi=1$
             are obtained from a linear extrapolation.}
\label{tab1}              
\end{table*}

The form factors $  \lambda_i (k, p)$, $i=1,2,3$, and $\tau_i (k, p)$, $i=3,5,8$, are proportional to the ghost-dressing function $G(q^2)$ defined by 
the propagator, 
\begin{equation}
  D^{a b} (q^2) = -\, \delta^{ab} \, \frac{G(q^2)}{q^2} \ , 
 \label{ghostprop} 
\end{equation} 
and renormalized as $G(\mu^2) = 1 $; $X_0^\xi (q^2)$ is the leading form factor of the quark-ghost scattering amplitude~\cite{Marciano:1977su,Davydychev:2000rt},
$H^{a\xi} (k, p) = H^\xi (k, p) t^a$, which can most generally be decomposed as,
\begin{align}
\hspace*{-2mm}
  H^\xi (k, p) & = \,  X_0^\xi ( k, p ) \mathbbm{1}_{D} + i X_1^\xi ( k, p )  \, \gamma\cdot k  \nonumber  \\
          & + \,  i  X_2^\xi ( k, p ) \  \gamma\cdot  p +  X_3^\xi ( k, p )\, \left[ \gamma \cdot k, \gamma \cdot p \, \right]   \, .
\label{Xidef1}       
\end{align}
Calculated in one-loop dressed approximation~\cite{Aguilar:2010cn,Rojas:2013tza}, \linebreak $X_0^\xi(k,p)$ can be projected out from the integral equation for the
quark-ghost scattering amplitude. We do so in the simplified kinematic configuration $k=-p=q/2$, thus omitting an angular dependence between $k$ and $p$, 
which is expressed by the following integral:
\begin{align}
 X_0^\xi (q^2)  & = \,   \tfrac{1}{4} \operatorname{Tr}_{\mathrm{CD}}\, H^\xi(q/ 2,-q / 2)            \nonumber \\ 
                 & =  \,  1 + \frac{C_A}{4} \, g^2 \! \int^\Lambda\!\!    \frac{d^4\ell}{(2\pi)^4}\ \Delta_{\mu \nu}^\xi (\ell)  D(\ell+q )   \nonumber \\ 
                  & \times  \,        \operatorname{Tr}_D \big [  G_{\nu}\,  S_\xi (\ell+\tfrac{q}{2})\, \Gamma_{\mu}^\xi ( \ell +\tfrac{q}{2}, - \tfrac{q}{2}) \big ] \ .
  \label{X0int}                          
\end{align}
Here, $C_A=3$ is the Casimir operator in the adjoint representation, 
\begin{equation}
  G_\nu=i (\ell_\nu+q_\nu ) H_1-i \ell_\nu H_2 \ ,
\label{ghostgluon}  
\end{equation}
is the dressed ghost-gluon vertex for which the form factor $H_1$ was calculated in Ref.~\cite{Dudal:2012zx}, $\ell$ is the gluon momentum exchanged between 
the quark and the ghost and in this configuration $q$ coincides with the gluon momentum in the DSE~\eqref{DSEquark}. The trace is over color and Dirac indices. 
Beyond Landau gauge, the integral in Eq.~\eqref{X0int} diverges since the gluon propagator~\eqref{gluonprop} is not  transverse anymore and does  therefore not 
project out the $\ell_\mu$ terms of the ghost-gluon vertex~\eqref{ghostgluon}. The divergence is mildly logarithmic and given that we employ lattice-QCD propagators 
for momenta $p \lesssim 8\,$GeV, no renormalization constant is necessary in the numerical integration, even for large $\Lambda$. Nonetheless, we multiply the left-hand 
side of Eq~\eqref{X0int} with a renormalization factor $Z_H$ and impose $X_0(\mu^2)=1$.

\begin{figure}[b!]
  \vspace{-2.5cm}
  \includegraphics[scale=0.98]{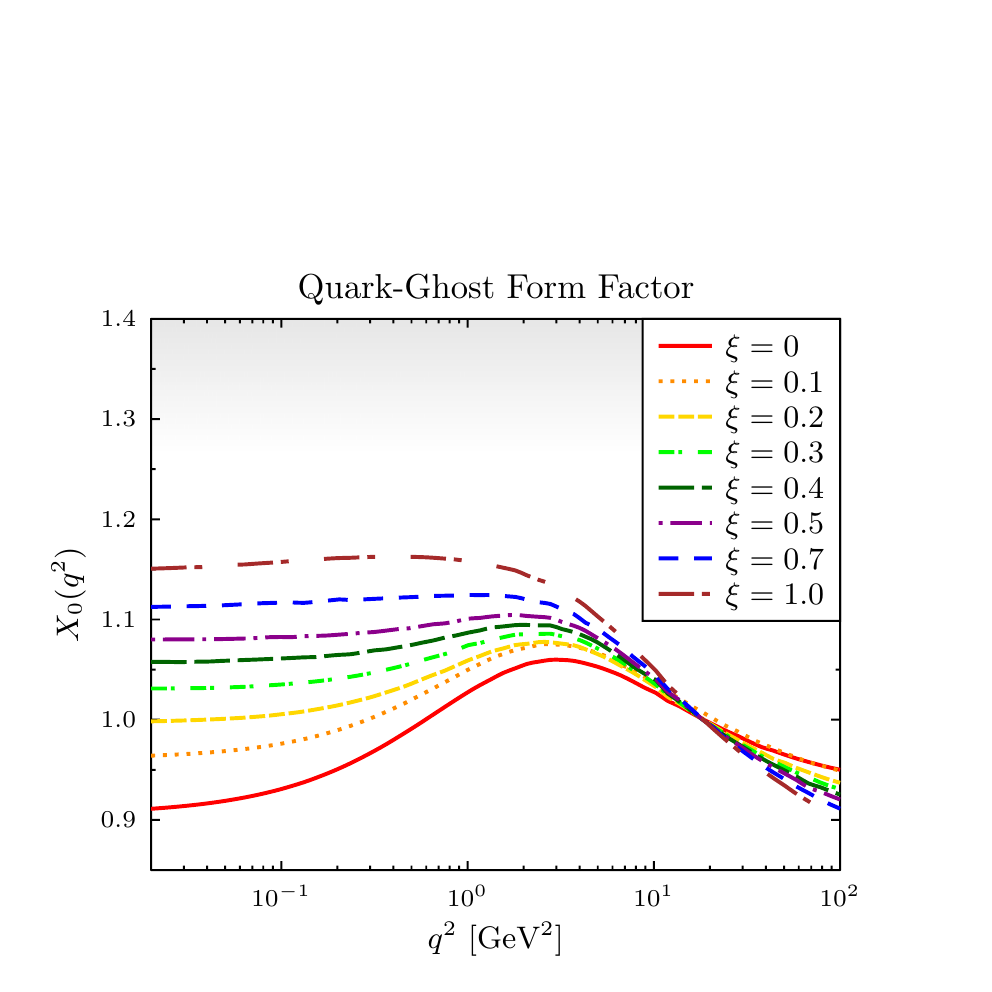}   
  \caption{Gauge-parameter  dependence of the form factor $X_0^\xi (q^2)$ associated with the quark-ghost scattering amplitude. }
 \label{figX0} 
\end{figure}

We neglect the remaining $X_i^\xi (k,p)$ form factors as they are suppressed with respect to $X_0^\xi (k, p)$~\cite{Aguilar:2016lbe} in Landau gauge. One may 
ask whether this approximation  is justified in other gauges as the form factors may vary in strength with the gauge parameter. This dependence is  depicted in 
Fig.~\ref{figX0} for $X_0^\xi (q^2)$ and discussed in  more detail in Section~\ref{sec3}. It turns out that setting $X_0^\xi (q^2) =1$ has only modest quantitative 
effects on $M_\xi(p^2)$ and  $Z_\xi(p^2)$, at least for  $\xi \in [0,1]$. Therefore, we make bold to assume that our observation is valid for the other form factors
and will analyze their behavior as a function of $\xi$ in a forthcoming study. We will see in Sec.~\ref{sec3} that the contribution of $X_0(q^2)$ to DCSB is small  
compared with the effect of the transverse vertex.

\begin{figure*}[t!]
\centering
  \includegraphics[scale=1]{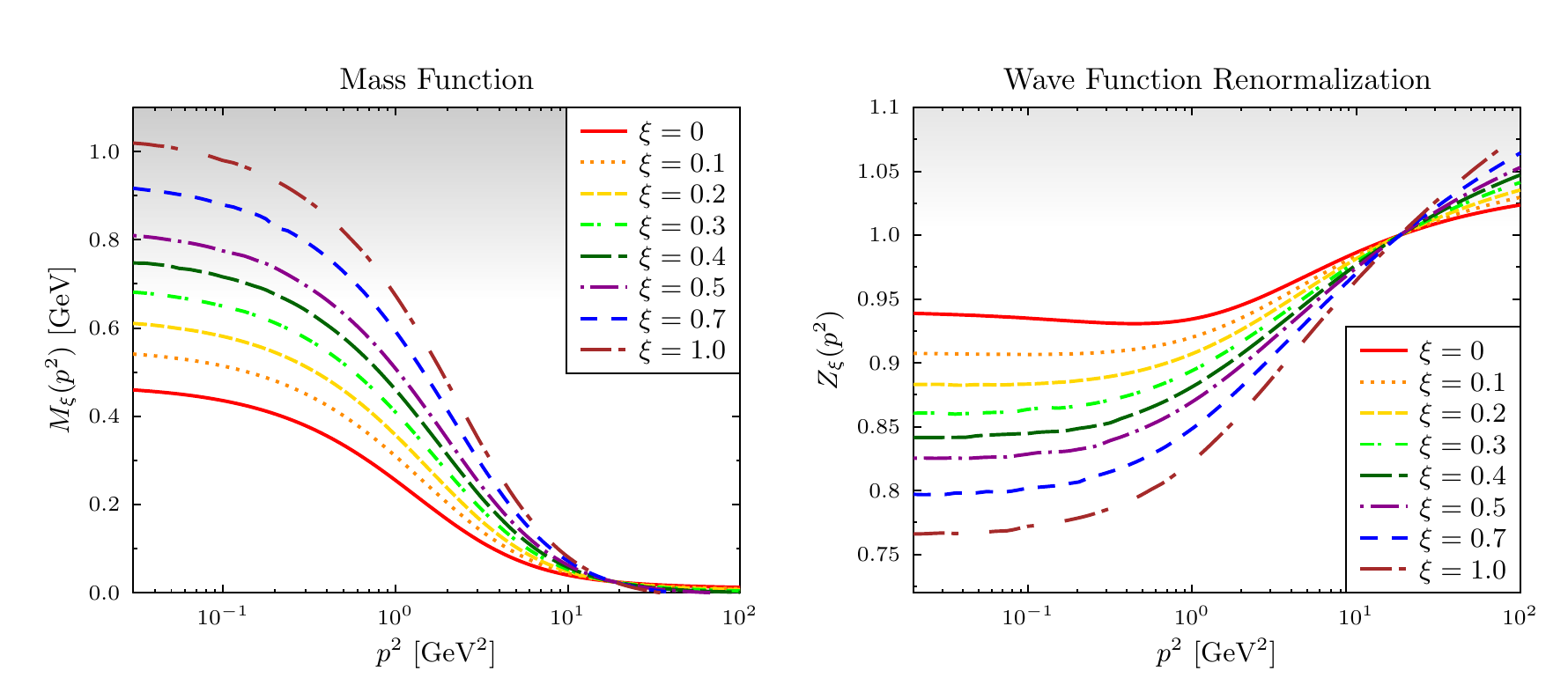}   \vspace*{-2mm}
  \caption{Mass function $M_\xi(p^2)$ and wave renormalization $Z_\xi(p^2)$ as functions of the gauge parameter $\xi$ obtained by solving the DSE~\eqref{finalDSEquark} 
               with the gluon and ghost propagators in Eqs.~\eqref{gluonparam} and \eqref{ghostparam}, respectively, and the vertex $\Gamma_\mu^\xi (k,p)$ of Eq.~\eqref{BallChiu} 
               with corresponding form factors $\lambda_i^\xi(k,p)$ and $\tau_i^\xi(k,p)$. }
 \label{figMZ-Rxi} 
\end{figure*} 

The $Y_i^{(A,S)}$ functions represent the form-factor decomposition of the Fourier transform of a four-point function in coordinate space in the TSTI. 
The latter involves a line integral over a nonlocal vector vertex and a Wilson line to preserve gauge invariance. As its matrix elements are rather complicated expressions, 
we refer to the discussion in Ref.~\cite{He:2009sj}. We note that the $Y_i^{(A,S)}$ functions are \emph{a priori\/} unknown and have been constrained by 
us~\cite{Albino:2018ncl}  with the vertex ansatz in Ref.~\cite{Bashir:2011dp}, i.e. we equated our transverse form factors~\eqref{tau1QCD}--\eqref{tau8QCD} with 
those of Ref.~\cite{Bashir:2011dp} and derived expressions for the $Y_i(k,p)$ that depend on $A(p^2)$, $B(p^2)$ and  $X_0(p^2)$. On the other hand, if the $Y_i^{(A,S)}$ 
form factors were known exactly, so would be the vertex. However,  additional input, e.g. from the  generalized LKFT that  describes the vertex itself, would be required 
for its full determination. We remind that we do not solve the BSE of the quark-gluon vertex, hence we do not  truncate the vertex form factors. Therefore, the expressions in 
Eqs.~\eqref{lambda1QCD} to \eqref{tau8QCD} represent a most complete form factor decomposition of the vertex that satisfies SU(3) gauge symmetry,  the $C$ and $P$ 
symmetries of the bare vertex and which is not plagued by kinematic singularities in the limit $k \to p$. 

In principle, one must also solve the gluon and ghost DSE which are coupled to the gap equation~\eqref{DSEquark}. We deliberately abstain from doing  so, as our interest is in 
employing quenched gluon and ghost propagators in $R_\xi$ gauge from lattice QCD~\cite{Bicudo:2015rma}. For practical reasons in their numerical implementation, 
we fit the gluon dressing functions with the parametrization~\cite{Dudal:2018cli}, 
\begin{align}
   \Delta_\xi (q^2 )  =  \frac{Z( q^{2}+M_{1}^{2})} {q^{4}+M_{2}^{2} q^{2}+M_{3}^{4}}
   \left [1 +  \omega \ln \left(\frac{q^{2}+M_0^2}{\Lambda_{\mathrm{QCD}}^2} \right ) \right ]^{-\gamma_\mathrm{gl}} \hspace*{-5mm} ,
 \label{gluonparam}   
\end{align}
where $\omega=11 N_c \,\alpha_{s}(\mu) /12 \pi$, $\Lambda_{\mathrm{QCD}}=0.425$~GeV and  $\gamma_\mathrm{gl} = (13-3\xi)/22$ is the 1-loop anomalous 
gluon dimension. The renormalization scale is $\mu =4.3$~GeV at which the strong coupling is chosen to be  $\alpha_{s} = 0.29$ in Landau gauge~\cite{Aguilar:2016ock}. 
This  parametrization is motivated by the refined Gribov-Zwanziger tree-level gluon propagator in the infrared domain and by the 1-loop renormalization group 
behavior for large momenta, which amounts to a renormalization-group improved Pad\'e approximation. We collect the values  $Z$, $M_0$, $M_1$, $M_2$ 
and  $M_3$ as a function of the gauge parameter $\xi$ in Table~\ref{tab1}. It turns out that all fit parameters depend linearly on $\xi$ and we take this 
feature to our advantage to extrapolate the parametrization to $\xi=1$.
%
\begin{figure}[b!]
  \vspace{-1.1cm}
\centering
  \includegraphics[scale=0.82]{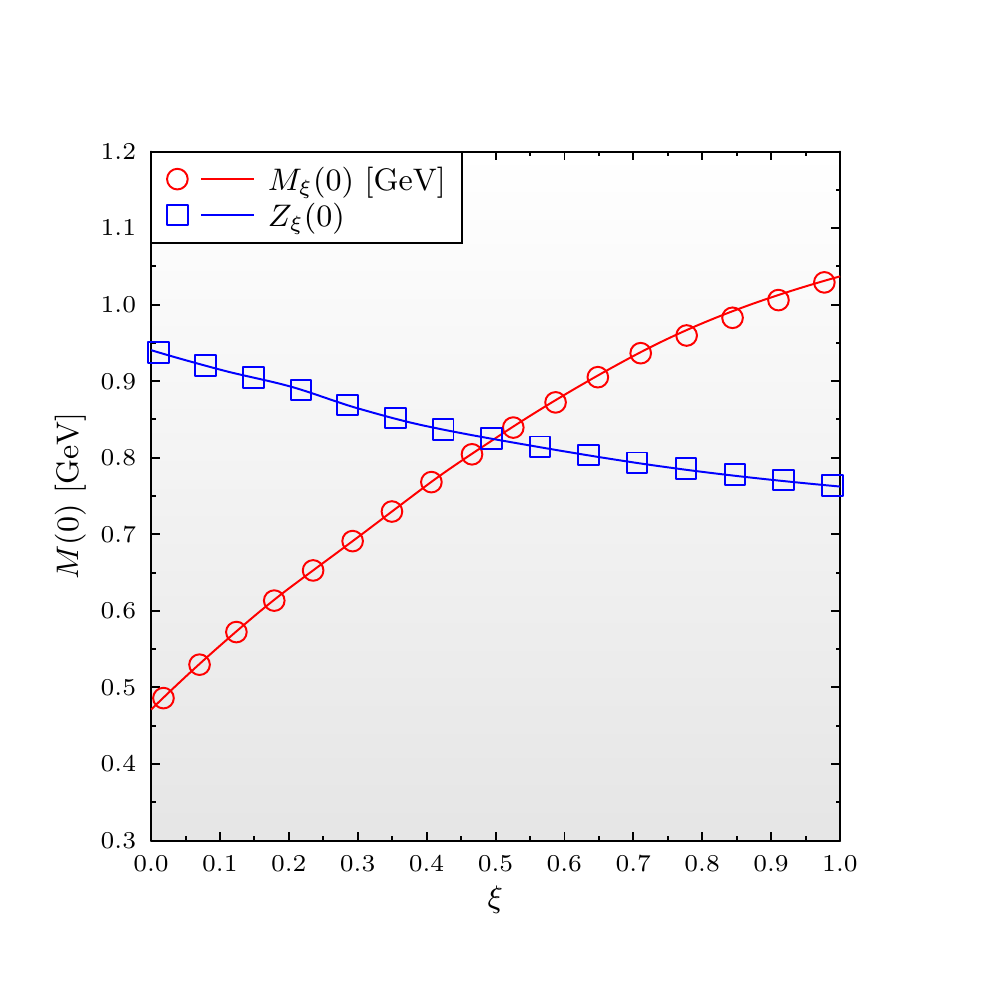}   
     \vspace{-0.9cm}
  \caption{Increasing and  decreasing  dependence of the mass function $M_\xi(0)$ and  wave renormalization $Z_\xi(0)$, respectively, on the gauge parameter $\xi$. }
 \label{figM0} 
\end{figure} 
%
\begin{figure*}[t!]
   \vspace*{-1cm}
\centering
  \includegraphics[scale=0.86]{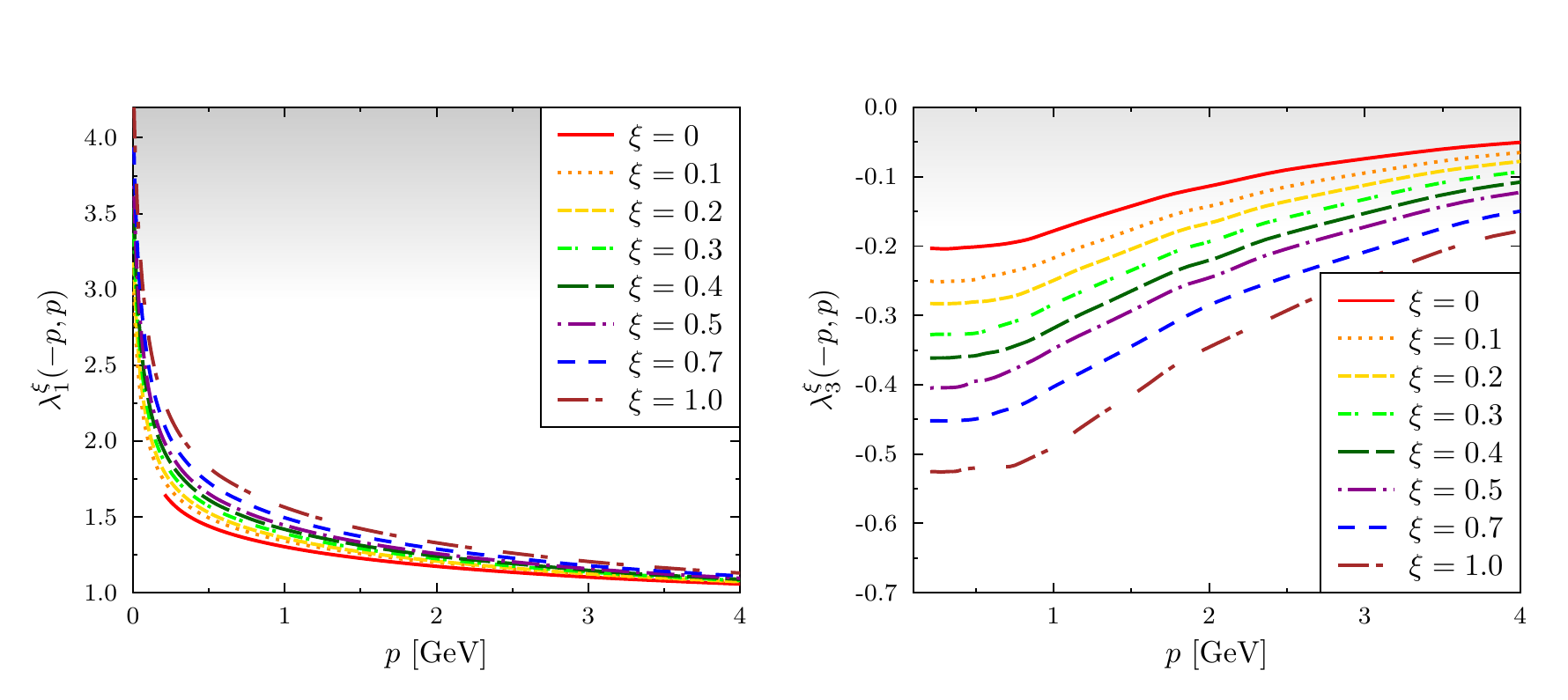}   \vspace*{-2mm}
  \caption{The form factors $\lambda_1^\xi (-p,p)$ and $\lambda_3^\xi (-p,p)$ as functions of the gauge parameter $\xi$, obtained with Eqs.~\eqref{lambda1QCD} and 
  \eqref{lambda3QCD} in the symmetric limit $k=-p$. Note that the dimensionless quantity $p\lambda_3^\xi$ is plotted. }
 \label{lambda13} 
\end{figure*} 

The bare lattice data~\cite{Duarte:2016iko} for the ghost propagator is parametrized with an analogous expression,
\begin{align}
    G  (q^2) &  =   \frac{Z(q^4+M_2^2 q^2+M_1^4)}{q^4+M_4^2 q^2+M_3^4} 
    \nonumber \\ & \times 
 \left [ 1 +\omega \ln \hspace{-1mm} 
 \left ( \frac{q^2+\frac{m_1^4}{q^2+m_0^2}}{\Lambda_\mathrm{QCD}^2} \right )   \right ]^{ \hspace{-1mm} \gamma_{\mathrm{gh} } } \hspace*{-2mm}  ,
  \label{ghostparam}   
\end{align}
independent of $\xi$, as supported by a preliminary study in lattice QCD~\cite{Cucchieri:2018leo}. The anomalous ghost dimension is \linebreak 
$\gamma_{\mathrm{gh}} = -9/44$, while $\omega$, $\Lambda_{\mathrm{QCD}}$ and $\mu$ are as in Eq.~\eqref{gluonparam}. \linebreak 
A least-squares fit yields $\chi^2/\mathrm{dof} = 0.247$ and the parameters: $Z =5.068 \pm 0.012$, $M_1^4 = 19.281 \pm 0.552 \,\mathrm{GeV}^4$, 
$M_2^2 = 27.721\pm 0.696 \,\mathrm{GeV}^2 $, $M_3^4  = 7.695\pm 0.329 \,\mathrm{GeV}^4$, $M_4^2 =  24.340\pm 0.565\, \mathrm{GeV}^2$, 
$m^2_0 = 0.527 \pm 1.263\,\mathrm{GeV}^2$, $m^4_1 = 0.018 \pm 0.035\,\mathrm{GeV}^4$; this fit must be normalized by a factor $N=1/4.706$
to ensure the renormalization condition $G(4.3\,\mathrm{GeV}) =1$.


 \section{Gauge dependence of the quark propagator and invariance of the quark condensate\label{sec3}}

With all the calculational tools and elements at hand in the gauge-parameter interval between the Landau and Feynman gauges, we can evaluate the gauge dependence 
of the quark's mass and wave-renormalization functions as well as that of the quark condensate.

\subsection{Gauge dependent mass and wave-renormalization functions}

Since the gauge propagators and the quark-gluon vertex are now determined we can proceed to solve the DSE~\eqref{DSEquark}. After taking the color trace,
and with the renormalization procedure detailed in Ref.~\cite{Albino:2021rvj}, the DSE becomes:
\begin{align}
\label{finalDSEquark}    
   S_\xi^{-1}(p) & = \, Z_2 \,  i\, \gamma\cdot p + Z_4 \, m (\mu  )  \nonumber \\ 
                      + \, Z_2 & \frac{16\pi \alpha_s^\xi}{3}   \!\int^\Lambda\!\!  \frac{d^4k}{(2\pi)^4}\  \Delta^\xi_{\mu\nu} (q)\, \gamma_\mu \, S_\xi(k)\,\Gamma_\nu^\xi (k,p) \ .                           
\end{align}
We plot the solutions of this DSE in Fig.~\ref{figMZ-Rxi} which clearly exposes that the mass function $M_\xi(p^2)$ increases with $\xi$  below the renormalization  
point $\mu = 4.3$~GeV, whereas the wave renormalization $Z_\xi(p^2)$ is suppressed. In fact, $M_\xi(0$) increases rather continuously with $\xi$ whereas $Z_\xi(0)$ 
slowly decreases as function of $\xi$, as can be read from Fig.~\ref{figM0}. We remark that  the gauge dependence is threefold in Eq.~\eqref{finalDSEquark}, 
namely $\xi$ enters directly via the longitudinal component and indirectly via the dressing function of the transverse component, $\Delta_\xi (q^2 )$, of the gluon 
propagator~\eqref{gluonprop}, but also via the gauge-dependent  strong coupling  $\alpha_s^\xi$ for which we use the parametrization~\cite{Aguilar:2016ock}:
\begin{equation}
    \alpha_s^\xi = 0.29 + 0.098 \xi - 0.064 \xi^2 \ .
  \label{strongcoupling}  
\end{equation}

The mass-function enhancement in Fig.~\ref{figMZ-Rxi} is due to the feedback of the gauge dependent $A_\xi (p^2)$ and $B_\xi (p^2) $ functions in the dressed 
quark-gluon vertex, namely Eqs.~\eqref{lambda1QCD} to \eqref{tau8QCD}, as $M_\xi(p^2)$ also increases when we keep the coupling $\alpha_s =0.29$ constant
for all values of $\xi$. Moreover, $M_\xi(p^2)$  slightly decreases with $\xi$ in the rainbow truncation, $\Gamma^\xi_\mu (k,p) = \gamma_\mu$, while the DSE solutions 
employing $\Gamma_\mu^{L\xi} (k,p)$ with Eqs.~\eqref{lambda1QCD} to \eqref{lambda4QCD} show an enhancement of the mass as a function of $\xi$. 
%
We emphasize that the overwhelming contribution to mass generation is due to the massive terms in the expressions for $\lambda_i^\xi(k,p)$ and $\tau_i^\xi(k,p)$.
While $X_0^\xi(q^2)$ is gauge dependent, as becomes clear from Fig.~\ref{figX0} where an enhancement of this form factor with increasing values of $\xi$ in 
the low-momentum region is observed, setting $X_0^\xi(q^2) = 1, \xi \in [0,1]$, leads to a 10\% reduction of $M_{\xi=1}(0)$ in Feynman gauge and to almost no
variation of $M_{\xi=0}(p^2)$ in Landau gauge. Neglecting $X_0^\xi(q^2)$ does not qualitatively alter our results, in particular our conclusions about the
quark condensate remain the same. 

We also illustrate the gauge-variation impact on the strength of the quark-gluon vertex in Fig.~\ref{lambda13} with the form factors $\lambda_1^\xi(k,p)$ and 
$\lambda_3^\xi (k,p)$ obtained in the symmetric limit $k=-p$ and with the solutions for $M_\xi(p^2)$ and $Z_\xi(p^2)$ in Fig.~\ref{figMZ-Rxi}. Both form factors are 
enhanced, though this is more so the case for $\lambda_3^\xi (-p,p)$. The latter is proportional to the mass function $B_\xi(p^2)$ whose variation with $\xi$ is more prominent.


\subsection{Quark condensate and gauge invariance}

\begin{figure}[t!]
\centering
  \includegraphics[scale=0.87]{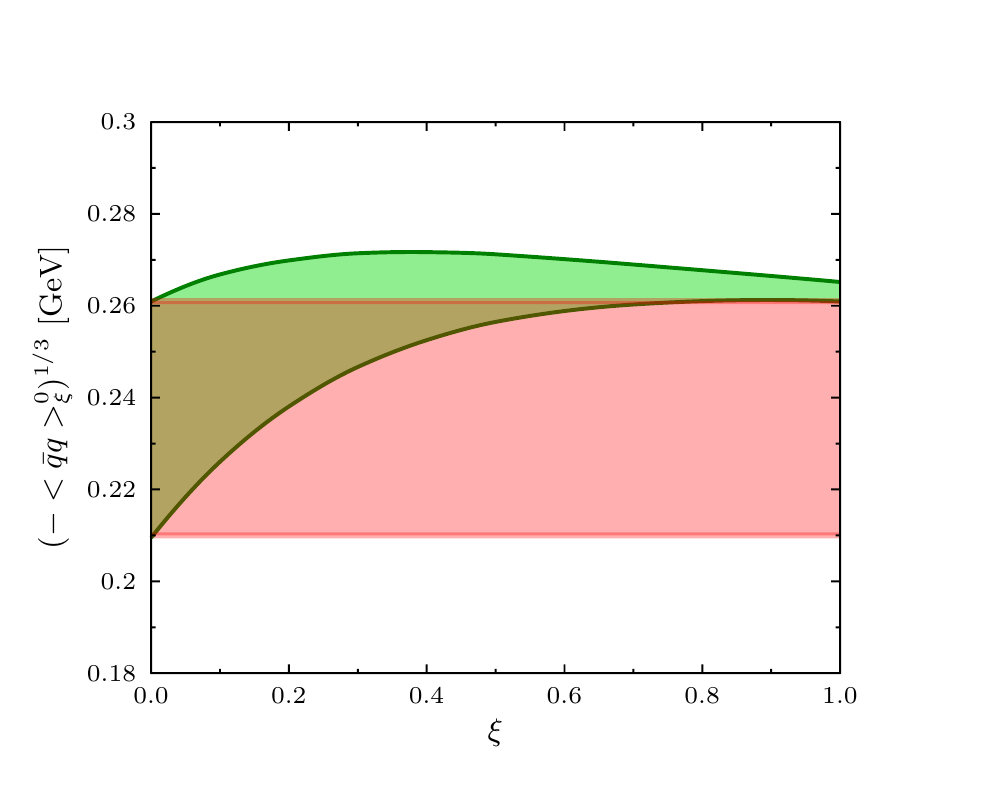}   
  \caption{Gauge dependence of the quark condensate. The horizontal pink-shaded band indicates the admissible region of a gauge-independent chiral 
                quark condensate as implied by the LKFT in QCD }
  \label{qqcondensate}
\end{figure}

As a practical application, we calculate the quark condensate which is an order parameter for DCSB and was shown to be a manifestly gauge-invariant 
quantity in any SU($N$) theory using the generalized LKFT (ABG) transformations~\cite{Aslam:2015nia}:
\begin{equation}
   -  \langle\bar{q} q \rangle^0_\xi  \, \equiv \, Z_4  N_c  \int^{\Lambda}  \!  \frac{d^4k}{(2\pi)^4} \, \operatorname{tr}_{D}  \left [ S_\xi^0 (k) \right ] \, .
\end{equation}
To this end, we obtain $M_\xi(p^2)$ and $Z_\xi(p^2)$ in the limit $m (\mu) \to 0$ with which we compute $ (-\langle\bar{q} q \rangle_\xi^0 )^{1/3}$ as a function of $\xi$. 
The result is presented in Fig.~\ref{qqcondensate}, where the central value exhibits a moderate dependence on $\xi$, i.e. a maximum deviation from the 
 Landau-gauge value of 7\% for $\xi \gtrsim 0.3$ after which its dependence on $\xi$ appears to become milder. The green-shaded band depicts 
 an error estimate due to the statistical error of  $\pm 10$\% of the gluon propagator.  This is because the lattice calculations of the gluon propagator in $R_\xi$ 
 gauges were performed in smaller physical volumes, $V\simeq (3.2\,\text{fm})^4$, and smaller gauge ensembles than for the ghost propagator for which the 
 fit reproduces the simulation results in a large physical volume, $V\simeq (8.1\, \text{fm})^4$, and with an ensemble of gauge configurations that is about three times 
 larger.  Therefore, the main error source is due to the gluon propagator and  we  neglect a statistical error associated with the ghost.

 Our error estimate only enters via the transverse part of the gluon propagator~\eqref{gluonprop}, yet with increasing values of the gauge parameter 
 the contribution of the longitudinal component is more important. This is the reason why the error band is wider in Landau gauge and narrows 
 towards Feynman gauge. We did not include the uncertainty of the strong coupling~\eqref{strongcoupling} and the systematic error due to 
our treatment of the nonlocal four-point functions in the TSTI.

\section{Final remarks}
\label{sec4}

We have calculated the quark propagator's dependence on the gauge parameter $\xi$ in covariant gauges. It turns out that the gluon  propagator parameterization 
and $M_\xi(0)$  can be fitted with linear functions of $\xi$ within the gauge-parameter interval, $\xi \in [0,0.5]$, while $Z_\xi(0)$ decreases in a proportionate manner.
Furthermore, the strength of the quark-gluon interaction increases with $\xi$ which illustrates the gauge dependence of the quark dynamics. Having said that, 
studying the local gauge transformation of the nonperturbative quark propagator directly from its solution in a given gauge is illuminating, as were similar QED3 
studies~\cite{Bashir:2005wt} that explicitly demonstrated the gauge independence of the fermion condensate. Both the Nielsen identities and the generalized LKFTs 
(ABG and MDSDB)~\cite{PhysRevD.101.085005,NIELSEN1975173,Aslam:2015nia,DeMeerleer:2018txc} formally provide the basis for this endeavor. However, it is 
not a straightforward exercise in QCD, even at the perturbative level~\cite{DallOlio:2021njq}.

While Green functions are not physical observables, they are essential objects for the elucidation and understanding of strong interactions. They enter each and 
every hadron observable computed from QCD's elementary degrees of freedom, namely quarks and gluons (and ghosts within covariant gauges). How the dependence 
on $\xi$ is washed out in the building of physical observables is a nontrivial problem that  constrains the truncated kernels of Bethe-Salpeter and Faddeev equations
and deserves further attention. The quark mass function inserted in a bound-state equation, conspires with  an intricate interaction kernel involving quarks and gluons at 
all energy scales in a manner that preserves axial vector Ward-Takahashi identity in addition to all other gauge identities and transformations mentioned in the text in 
detail, yielding gauge-independent physical observables. Any variation of these observables with the gauge parameter is a measure of our departure from the full 
implementation of the generalized LKFT and the Nielsen identities.

Our analysis of the DSE relies on the dressed vertex~\eqref{BallChiu} described by form factors $\lambda_i^{\xi} (k,p)$ and $\tau_i^{\xi} (k,p)$, which are plotted 
for $\xi=0$ in Ref.~\cite{El-Bennich:2022obe} and show to be in reasonable agreement with those obtained in lattice QCD simulations~\cite{Kizilersu:2021jen}. 
We are thus encouraged by our exploratory study on the gauge dependence of the quark propagator, as they provide a valuable initial step in the numerical 
representation of the generalized  LKFT and Nielsen identities in QCD despite the current limitations. However, further efforts must be made to include all
$X_i^\xi (k,p)$ form factors of the quark-ghost scattering amplitude along with their angular dependence, to derive a true analytic expression for the four-point 
function in the TSTI and to explore the dependence on these hitherto ignored features in the current analysis.


\section*{Acknowledgements}

The work of A.B. is supported in part by the US Department of Energy (DOE) Contract no.~DE-AC05-06OR23177, under which Jefferson Science Associates, 
LLC operates Jefferson Lab. It received funds from the {\em Coordinaci\'on de la Investigaci\'on Cient\'ifica} (CIC) of the University of Michoac\'an through grant 
no.~4.10. B.E. receives financial support from the Brazilian agencies FAPESP, grant no.~2018/20218-4, and CNPq, grant no.~428003/2018-4.  F.E.S. is a CAPES-PNPD 
postdoctoral  fellow, grant no.~88882.314890/2013-01. J.R.S. was supported by a CAPES PhD fellowship.  O.O. acknowledges support from FCT, Portugal, 
project nos. UID/FIS/04564/2019 and  UID/FIS/04564/2020. This work is  part of the  network project INCT-F\'isica Nuclear e Aplica\c{c}\~oes, no. 464898/2014-5.


\bibliography{references}

\end{document}